\documentclass{ws-mpla}

\begin{document}

\markboth{Morozova, Ahmedov, Abdujabbarov  $\&$ Mamadjanov}
{{Quantum Interference Effects Around Compact Objects in
Braneworld}}

%%%%%%%%%%%%%%%%%%%%% Publisher's Area please ignore %%%%%%%%%%%%%%
\catchline{}{}{}{}{}
%%%%%%%%%%%%%%%%%%%%%%%%%%%%%%%%%%%%%%%%%%%%%%%%%%%%%%%%%%%%%%%%%%%

\title{Plasma Magnetosphere of Rotating Magnetized Neutron Star in
the Braneworld}

\author{\footnotesize Morozova V.S.}

\address{Institute of Nuclear Physics, Ulughbek, Tashkent 100214, Uzbekistan\\
    Ulugh Beg Astronomical Institute, Astronomicheskaya 33,
    Tashkent 100052, Uzbekistan \\}

\author{\footnotesize Ahmedov B.J.}

\address{Institute of Nuclear Physics, Ulughbek, Tashkent 100214, Uzbekistan\\
    Ulugh Beg Astronomical Institute, Astronomicheskaya 33,
    Tashkent 100052, Uzbekistan \\
    ahmedov@astrin.uzsci.net}

    \author{\footnotesize Abdujabbarov A.A.}

\address{Institute of Nuclear Physics, Ulughbek, Tashkent 100214, Uzbekistan\\
    Ulugh Beg Astronomical Institute, Astronomicheskaya 33,
    Tashkent 100052, Uzbekistan \\
    ahmadjon@astrin.uzsci.net}

\author{\footnotesize Mamadjanov A.I.}

\address{Institute of Nuclear Physics, Ulughbek, Tashkent 100214, Uzbekistan\\
    Ulugh Beg Astronomical Institute, Astronomicheskaya 33,
    Tashkent 100052, Uzbekistan \\
mamadjanov@inp.uz}

\maketitle

\pub{Received (Day Month Year)}{Revised (Day Month Year)}

\begin{abstract}

Plasma magnetosphere surrounding rotating magnetized neutron star
in the braneworld has been studied. For the simplicity of
calculations Goldreich-Julian charge density is analyzed for the
aligned neutron star with zero inclination between magnetic field
and rotation axis. From the system of Maxwell equations in
spacetime of slowly rotating star in braneworld, second-order
differential equation for electrostatic potential is derived.
Analytical solution of this equation indicates the general
relativistic modification of an accelerating electric field and
charge density along the open field lines by brane tension. The
implication of this effect to the magnetospheric energy loss
problem is underlined. It was found that for initially zero
potential and field on the surface of a neutron star, the
amplitude of the plasma mode created by Goldreich-Julian charge
density will increase in the presence of the negative brane
charge. Finally we derive the equations of motion of test
particles in magnetosphere of slowly rotating star in the
braneworld. Then we analyze particle motion in the polar cap and
show that brane tension can significantly change conditions for
particle acceleration in the polar cap region of the neutron star.

\end{abstract}

%\ccode{PACS Nos.: 04.50.-h, 04.40.Dg, 97.60.Gb.}

\newpage

\section{Introduction}
\label{intro}

Amazing idea about hidden extra-dimensions of our Universe
attracted scientific interest almost from the beginning of 20th
century. First attempts of building multidimensional models were
proposed by Kaluza in order to unify electromagnetism with gravity
already in 1920s \cite{k21}. Then these ideas found reflection in
elegant string theory which is the subject of extensive research
in modern physics and promise to throw light upon many puzzles of
nature. One of the most recent theories including extra dimensions
is the braneworld picture of the Universe.

The braneworld model was first proposed by \cite{RaSu99} assuming
that our four-dimensional space-time is just a slice of
five-dimensional bulk. According to this model only gravity is the
force which can freely propagate between our space-time and bulk
while other fields are confined to four-dimensional Universe. In
this view it is noteworthy to look for effects of fifth dimension
on our world in frame of theory of gravity i.e. general
relativity. Possible tools for proving braneworld model should be
found from astrophysical objects, namely, compact objects, for
which effects of general relativity are especially strong. For
example, investigations of cosmological and astrophysical
implications of the braneworld theories may be found in
\cite{maar00}, \cite{cs01}, \cite{lan01}, \cite{hm03},
\cite{ger06}, \cite{kg08}, \cite{MaMu05}. Review of braneworld
models is given e.g. in \cite{maar04}. In the context of the
braneworld, a method to find consistent solutions to Einstein's
field equations in the interior of a spherically symmetric, static
and non uniform stellar distribution with Weyl stresses is
developed in the work of \cite{ovalle}.

For astrophysical interests, static and spherically symmetric
exterior vacuum solutions of the braneworld models were initially
proposed by Dadhich et al. \cite{dmp00} which have the
mathematical form of the Reissner-Nordstr\"{o}m solution, in which
a tidal Weyl parameter $Q^\ast$ plays the role of the electric
charge squared of the general relativistic solution.

Observational possibilities of testing the braneworld black hole
models at an astrophysical scale have been intensively discussed
in the literature during the last several years, for example,
through the gravitational lensing, the motion of test particles,
and the classical tests of general relativity (perihelion
precession, deflection of light, and the radar echo delay) in the
Solar System (see \cite{lobo08}). In paper \cite{pkh08} the energy
flux, the emission spectrum, and accretion efficiency from the
accretion disks around several classes of static and rotating
braneworld black holes have been obtained. The complete set of
analytical solutions of the geodesic equation of massive test
particles in higher dimensional spacetimes which can be applied to
braneworld models is provided in the recent paper \cite{Lam08}.
The relativistic quantum interference effects in the spacetime of
slowly rotating object in braneworld and phase shift effect of
interfering particle in neutron interferometer have been studied
in the recent paper \cite{mht10}.
The influence of the tidal charge onto profiled spectral lines
generated by radiating tori orbiting in the vicinity of a rotating
black hole has been studied in paper \cite{sstuch09}. Authors
showed that with lowering the negative tidal charge of the black
hole, the profiled line becomes flatter and wider, keeping their
standard character with flux stronger at the blue edge of the
profiled line. The role of the tidal charge in the orbital
resonance model of quasiperiodic oscillations in black hole
systems has been investigated in paper \cite{stuch09}. The
influence of the tidal charge parameter of the braneworld models
on some optical phenomena in rotating black hole spacetimes has
been extensively studied in paper \cite{ssstuch09}.

A braneworld corrections to the charged rotating black holes and
to the perturbations in the electromagnetic potential around black
holes are studied, for example, in \cite{aliev05,aa10}.  The
motion of test particles near black holes immersed in an
asymptotically uniform magnetic field and some gravity surrounding
structure, which provides the magnetic field has been intensively
studied in paper \cite{kon06}. The author has calculated the
binding energy for spinning particles on circular orbits. The
bound states of the massive scalar field around a rotating black
hole immersed in the asymptotically uniform magnetic field are
considered in paper \cite{kon07}. The uniform magnetic field in
the background of a five dimensional black hole has been
extensively studied in \cite{alfr04}. In particular, authors
presented exact expressions for two forms of an electromagnetic
tensor and the electrostatic potential difference between the
event horizon of a five dimensional black hole and the infinity.

Among astrophysical objects which can be useful in investigating
high-dimensional models particular place belongs to radio pulsars.
Radio pulsars are rotating highly magnetized neutron stars,
producing radio emission above the small area of its surface
called polar cap. \cite{gj69} proved that such a rotating highly
magnetized star cannot be surrounded by vacuum due to generation
of strong electric field pulling out charged particles from the
surface of the star. They proposed first model of the pulsar
magnetosphere containing two distinct regions: the region of
closed magnetic field lines, where plasma corotates with the star
as a solid body, and the region of open magnetic field lines,
where radial electric field is not completely screened with plasma
particles and plasma may leave the neutron star along magnetic
field lines. Radio emission is generated due to continuous cascade
generation of electron-positron pairs in the magnetosphere above
the polar cap. Thorough research on structure and physical
processes in pulsar magnetosphere can be found in works of
\cite{gj69}, \cite{s71}, \cite{m71}, \cite{rs75}, \cite{as79},
\cite{MusHar97}. Although a self-consistent pulsar magnetosphere
theory is yet to be developed, the analysis of plasma modes in the
pulsar magnetosphere based on the above-mentioned papers provides
firm ground for the construction of such a model.

It was shown by a number of authors that effects of general
relativity play very important role in physics of pulsars. General
relativistic effects on the vacuum electromagnetic fields around
slowly rotating magnetized neutron stars were investigated by
\cite{r1}, \cite{r2}. The effect of general relativistic frame
dragging effect in the plasma magnetosphere was investigated in
\cite{bes90}, \cite{MusTsy92}, \cite{MusHar97}, \cite{ma00},
\cite{bes05} and proved to be crucial for the conditions of
particle acceleration in the magnetosphere and, therefore, for
generation of radio emission. From the above mentioned papers it
is seen that the general relativistic effects in the magnetosphere
of pulsars are not negligible and should be carefully considered.

Our preceding paper \cite{af08} was devoted to the stellar
magnetic field configurations of relativistic stars in dependence
on brane tension and the present research extends it to the case
of rotating relativistic star. Here we will consider rotating
spherically symmetric star in the braneworld endowed with strong
magnetic fields. We assume that the star has dipolar magnetic
field and the field energy is not strong enough to affect the
spacetime geometry, so we consider the effects of the
gravitational field of the star in the braneworld on the magnetic
and electric field structure without feedback. That is our present
research is devoted to the strong gravity effects in pulsar
magnetosphere in frame of the braneworld model. We use results of
the work of \cite{MusTsy92} initiated the detailed general
relativistic derivation of magnetospheric electromagnetic fields
around rotating magnetized neutron star. In section 2 we analyse
Goldreich-Julian charge density for the case of slowly rotating
neutron star in the braneworld and get solution for Poisson
equation in this case. In sections 3 and 4 we have studied plasma
modes along the open field lines of the rotating magnetized
neutron star in the braneworld. Section~\ref{acceleration} is
devoted to the charged particle acceleration in the polar cap of
the rotating star in the braneworld. Section~\ref{concl}
summarizes the obtained results.

Throughout the paper we use a space-like signature $(-,+,+,+)$ and
a system of units in which $G = 1 = c$. Latin indices run $1,2,3$
and Greek ones from $0$ to $3$.

\section{Plasma Magnetosphere of Slowly Rotating Magnetized
Star in the Braneworld} \label{ms}

In a pioneering work, \cite{gj69} have shown that a strongly
magnetized, highly conducting neutron star, rotating about the
magnetic axis, would spontaneously build up a charged
magnetosphere. Strong magnetic field of the star together with
rotation create strong radial electric field component
$E_{\parallel}\sim 10^{10}-10^{12}\ V\ cm^{-1}$ on the surface of
the star. Such a strong field make charged particles escape from
the surface of the star and form plasma magnetosphere around the
star. Plasma charges, in turn, partially screen radial electric
field and ${\bf{E}}\times{\bf{B}}$ drift sets them into corotation
with the star. The magnetosphere charge density which would be
necessary for complete screening of $E_{\parallel}$ is called the
corotation charge density or the Goldreich-Julian density. In the
region of the polar cup magnetospheric charge does not screen
$E_{\parallel}$ completely, what results in continuous flow of
accelerated charged particles from the surface of the polar cap
responsible for later generation of radioemission from the region.

In general, Goldreich-Julian charge density $\rho_{GJ}$ can be
found through the formula
\begin{equation}
\label{rhoGJdef} \rho_{GJ}=-\frac{1}{4\pi}{\mathbf\nabla}(N{\bf
g}\times{\bf B})\ ,
\end{equation}
where $g_{i}=-g_{0i}/g_{00}$ can be found with the help of the
metric of given space-time, $N$ is the lapse function.

The metric, describing external space-time for the rotating star
in the braneworld, has the following form in spherical coordinates
\begin{eqnarray}
&& ds^2=-A^2dt^2+H^2dr^2+r^2d\theta^2+r^2\sin^2\theta
d\phi^2\nonumber\\&&\qquad \quad -
2\tilde{\omega}(r)r^2\sin^2\theta dt d\phi\ .\label{metric}
\end{eqnarray}

This metric is derived from the general solution for the rotating
star on the branes \cite{pkh08} under the assumption that specific
angular momentum  $a=J/M$ is small. In the equation (\ref{metric})
\begin{equation}
A^2(r)\equiv\left(1-\frac{2M}{r}+\frac{Q^\ast}{r^2}\right)=H^{-2}(r),\
\ \ \ \ \ \ \ r>R
\end{equation}
and  $\tilde{\omega}(r)=\omega(1-Q^\ast/2rM)$, $Q^\ast$ is the
bulk tidal charge, $M$ is the mass of the star, $\omega=2Ma/r^3$
is the angular velocity of the dragging of inertial frames.

In frames of metric (\ref{metric}) one can obtain the following
equations for the vector  ${\bf g}$
\begin{equation}
\label{g} {\bf
g}=\frac{1}{A^2}\left[{\mathbf\Omega}-\left(1-\frac{Q^\ast}{2rM}\right)
{\mathbf\omega}\right]\times{\bf r}
\end{equation}

and Goldreich-Julian charge density
\begin{equation}
\label{rhoGJcont} \rho_{GJ}=-\frac{1}{4
\pi}{\mathbf\nabla}\left\{\frac{1}{A}\left[1-\left(1-\frac{Q^\ast}{2\eta
RM}\right)\frac{\kappa}{ \eta^3}\right] {\bf u}\times{\bf B}
\right\}\ ,
\end{equation}
where ${\bf u}={\mathbf\Omega}\times{\bf r}$, $\eta=r/R$ is the
dimensionless radial coordinate, parameter
$\kappa\equiv\varepsilon\beta$, $\varepsilon=2M/R$ is the
compactness parameter, and $\beta=I/I_0$ is the stellar moment of
inertia in the units of $I_0=MR^2$.

%дипол€рна€ конфигураци€ дл€ магнитного пол€ - почему используем

In the work \cite{MusTsy92} the expressions for the components of
dipolar magnetic field ${\bf B}$  in the vicinity of the slowly
rotating neutron star are presented. In these calculations the
metric of space-time was assumed to have the following form:
\begin{eqnarray}
&& ds^2=-N^2dt^2+N^{-2}dr^2+r^2d\theta^2+r^2\sin^2\theta
d\phi^2\nonumber \\ && \qquad\quad- 2\omega r^2\sin^2\theta dt
d\phi\ , \label{metricN}
\end{eqnarray}
where $N^2\equiv(1-2M/r)$. Being measured by zero angular momentum
observer (ZAMO) with 4-velocity $u_\alpha =
\left\{-N,0,0,0\right\}$ components of dipolar magnetic field have
the form
\begin{eqnarray}
&& B^{\hat r}=B_0\frac{f(\eta)}{f(1)}\eta^{-3}\cos\theta\ ,
\label{B} \\&& B^{\hat\theta} =\frac{1}{2} B_0
N\left[-2\frac{f(\eta)}{f(1)}+
\frac{3}{(1-\varepsilon/\eta)f(1)}\right]\eta^{-3}\sin\theta\ ,
\nonumber
\end{eqnarray}
with
\begin{equation}
\label{f}
f(\eta)=-3\left(\frac{\eta}{\varepsilon}\right)^3\left[\ln
\left(1-\frac{\varepsilon}{\eta}\right)+\frac{\varepsilon}{\eta}
\left(1+\frac{\varepsilon}{2\eta}\right)\right]\ ,
\end{equation}
where $B_0\equiv 2\mu/R^3$ is the Newtonian value of the magnetic
field at the pole of star, hats label the orthonormal components
and $\mu$ is the magnetic moment. In the spacetime of the slowly
rotating neutron star in the braneworld Maxwell equations for the
magnetic form have a little more complicated form and avoid exact
analytical solution yet. Nevertheless, one may expect that
corrections to the magnetic field due to tidal charge $Q^\ast$
will be small in comparison with main expression. In our research
for the simplicity of the calculations we are going to neglect
these corrections and use equation~(\ref{B}) for the magnetic
field of the star.

Inserting equation (\ref{B}) into equation (\ref{rhoGJcont}) we
get the following final form for the Goldreich-Julian charge
density
\begin{equation}
\label{rhoGJ} \rho_{GJ}=-\frac{\Omega B_0}{2\pi c
}\frac{1}{A\eta^3}\frac{f(\eta)}{f(1)}\left\{1-
\left(1-\frac{Q^\ast}{2\eta
RM}\right)\frac{\kappa}{\eta^3}\right\}\ .
\end{equation}

In the papers \cite{MusTsy91}, \cite{MusTsy92} the expression for
the polar angle $\Theta$ of the last open magnetic line as a
function of $\eta$  can be found
\begin{eqnarray}
\label{Theta} &&
\Theta\cong\sin^{-1}\left\{\left[\eta\frac{f(1)}{f(\eta)}\right]^{1/2}
\sin\Theta_0\right\}\ ,\nonumber\\&&
\Theta_0=\sin^{-1}\left(\frac{R}{R_{LC}f(1)}\right)^{1/2}\ ,
\end{eqnarray}
where  $\Theta_0$ is the magnetic colatitude of the last open
magnetic line at the stellar surface, $R_{LC}=c/\Omega$ is the
light-cylinder radius.

Assuming the magnetic field of a neutron star to be stationary in
the corotating frame, from the system of Maxwell equations the
following Poisson equation for the scalar potential $\Phi$ can be
derived (look \cite{MusTsy92})
\begin{equation}
\label{Poiss}
{\mathbf\nabla}\cdot\left(\frac{1}{A}{\mathbf\nabla}\Phi\right)=
-4\pi(\rho-\rho_{GJ})\ ,
\end{equation}
where $\rho-\rho_{GJ}$ is the effective space charge density being
responsible for production of unscreened parallel electric field.

Choosing the following form for the charge density $\rho$ in the
vicinity of the surface of the neutron star
\begin{equation}
\label{rho} \rho=\frac{\Omega B_0}{2\pi c
}\frac{1}{A\eta^3}\frac{f(\eta)}{f(1)}C(\xi)\ ,
\end{equation}
where $\xi=\theta/\Theta$ is the dimensionless angular variable,
$C(\xi)$ is an unknown angular dependent function to be defined
from the boundary conditions, and inserting equations (\ref{rho})
and (\ref{rhoGJ}) into the Poisson equation (\ref{Poiss}) one can
obtain under the approximation of small angles $\theta$ the
following differential equation
\begin{eqnarray}
&&\label{diff1}
R^{-2}\bigg\{A\frac{1}{\eta^2}\frac{\partial}{\partial\eta}\left(\eta^2
\frac{\partial}{\partial\eta}\right)\\
&&\qquad\qquad+\frac{1}{A\eta^2\theta}\left[
\frac{\partial}{\partial\theta}\left(\theta
\frac{\partial}{\partial\theta}\right)+\frac{1}{\theta}
\frac{\partial^2}{\partial\phi^2}\right]\bigg\}\Phi= \nonumber
\\&&\hspace{-0.5cm}
-4\pi\frac{\Omega B_0}{2\pi
c}\frac{1}{A\eta^3}\frac{f(\eta)}{f(1)}\left\{1-\left(1-\frac{Q^\ast}{2\eta
R M}\right)\frac{\kappa}{\eta^3}+C(\xi)\right\}. \nonumber
\end{eqnarray}

Our further discourses are based on the extension of
work~\cite{MusTsy92}. Using dimensionless function
$F=\eta\Phi/\Phi_0$, where $\Phi_0=\Omega B_0 R^2$ and variables
$\eta$ and $\xi$, one can rewrite the equation (\ref{diff1}) as
\begin{eqnarray}
&&\hspace{-0.5cm}
\left[\frac{d^2}{d\eta^2}+\Lambda^2(\eta)\frac{1}{\xi}\frac{\partial}
{\partial\xi}
\left(\xi\frac{\partial}{\partial\xi}\right)\right]F=- \frac{2}{
%\left(
\eta^2-\varepsilon\eta+\tilde{Q^\ast}
%\right)
} \nonumber\\ &&\frac{f(\eta)}{f(1)}
\left[1-\left(1-\frac{Q^\ast}{2\eta R
M}\right)\frac{\kappa}{\eta^3}+C(\xi)\right]\ ,\label{difLam}
\end{eqnarray}
where
$\Lambda(\eta)=[\eta\Theta(\eta)(1-\varepsilon/\eta+\tilde{Q^\ast}/\eta^{2})^{1/2}]^{-1}$
 and $\tilde{Q^\ast}=Q^\ast/R^{2}$.

One can apply Fourier-Bessel transformation
\begin{eqnarray}
&& F(\eta,\xi)=\sum_{i=1}^{\infty}F_i(\eta)J_0(k_i\xi)\ ,
\nonumber\\&&  F_i(\eta)=\frac{2}{[J_1(k_i)]^2}\int^1_0\xi
F(\eta,\xi)J_0 (k_i\xi)d\xi\ , \label{FourBes}
\end{eqnarray}
with the relation
\begin{equation}
\sum_{i=1}^{\infty}\frac{2}{k_i J_1(k_i)}J_0(k_i\xi)=1\ ,
\end{equation}
to obtain equation (\ref{difLam}) in the form
\begin{eqnarray}
&& \left(\frac{d^2}{d\eta^2}-\gamma^2_i(\eta)\right)F_i=-
\frac{2}{\eta^2-\varepsilon\eta+\tilde{Q^\ast}}\frac{f(\eta)}{f(1)}
\nonumber
\\&&
\left[\frac{2}{k_i J_1(k_i)}\left\{1-\left(1-\frac{Q^\ast}{2\eta R
M}\right)\frac{\kappa}{\eta^3}\right\}+ C_i\right]
,\label{difgamma}
\end{eqnarray}

where $\gamma^2_i=k^2_i\Lambda^2$, $k_i$ are positive zeros of the
functions $J_0$.

Considering a region near the surface of the star, where
$z=\eta-1\ll1$, and using following boundary conditions (that is
the conditions of equipotentiality of the stellar surface and zero
steady state electric field at $r=R$)
\begin{equation}
F_i|_{z=0}=0\ ,\ \frac{\partial F_i}{\partial z}|_{z=0}=0\
\end{equation}
one can find the expression for the scalar potential $\Phi$ near
the surface of the star and corresponding to this potential
component of the electric field $E_{\|}$, being parallel to the
magnetic field (see the discourses of~\cite{MusTsy92} work):
\begin{eqnarray}
&&
\Phi=\frac{12\Phi_0}{\eta}\sqrt{1-\varepsilon+\tilde{Q^\ast}}\kappa
\left(1-\frac{2Q^\ast}{3M R}\right) \Theta^3_0\nonumber
\\&& \sum^{\infty}_{i=1}\Bigg[\exp\bigg\{\frac{k_i(1-\eta)}
{\Theta_0\sqrt{1-\varepsilon+\tilde{Q^\ast}}}\bigg\}-1\nonumber
\\&&\qquad\qquad +\frac{k_i(\eta-1)}
{\Theta_0\sqrt{1-\varepsilon+\tilde{Q^\ast}}}
\Bigg]\frac{J_0(k_i\xi)}{k_i^4J_1(k_i)}\ ,
\end{eqnarray}
\begin{eqnarray}
&& E_{\|}=-\frac{12\Phi_0}{R}\kappa\left(1-\frac{2Q^\ast}{3M
R}\right)\Theta^2_0 \nonumber
\\&& \sum^{\infty}_{i=1}\left[1-\exp\left\{\frac{k_i(1-\eta)}
{\Theta_0\sqrt{1-\varepsilon+\tilde{Q^\ast}}}\right\}\right]
\frac{J_0(k_i\xi)}{k_i^3J_1(k_i)}\ .
\end{eqnarray}

Considering now the region $\Theta_0\ll\eta-1\ll R_{LC}/R$, where
$|d^2F_i/d\eta^2|\ll\gamma^2_i(\eta)|F_i|$ one can see that
equation (\ref{difgamma}) becomes
\begin{eqnarray}
&& -\gamma^2_i(\eta)F_i=
-\frac{2}{\eta^2-\varepsilon\eta+\tilde{Q^\ast}}
\frac{f(\eta)}{f(1)} \nonumber
\\&& \left[\frac{2}{k_i J_1(k_i)}\left\{1-\left(1-\frac{Q^\ast}{2\eta R
M}\right)\frac{\kappa}{\eta^3}\right\}+C_i\right]\ ,
\end{eqnarray}
from which it immediately follows
\begin{eqnarray}
&& \hspace{-0.6cm}
F_i=4\kappa\theta^2(\eta)\frac{f(\eta)}{f(1)}\bigg[1-\frac{Q^\ast}{2R
M}-\left(1-\frac{Q^\ast}{2\eta R M}\right)\frac{1}{\eta^3} \nonumber\\
&& -\frac{3}{\gamma_i(1)}\left(1-\frac{2Q^\ast}{3M
R}\right)\bigg]\frac{1}{k_i^3 J_1(k_i)}\simeq \nonumber \\ &&
\hspace{-0.6cm}  4\kappa\Theta_0^2\eta\left[1-\frac{Q^\ast}{2R
M}-\left(1-\frac{Q^\ast}{2\eta R
M}\right)\frac{1}{\eta^3}\right]\frac{1}{k_i^3 J_1(k_i)}  .
\end{eqnarray}

Using this expression for $F_i$ one can obtain the scalar
potential in the region at distances greater than the polar cap
size as
\begin{eqnarray}
&& \hspace{-0.7cm} \Phi=\frac{\Phi_0}{\eta}F=2\Phi_0\Theta^2_0\kappa \nonumber \\
&& \left[1-\frac{Q^\ast}{2R M}-\left(1-\frac{Q^\ast}{2\eta R
M}\right)\frac{1}{\eta^3}\right]\sum_i\frac{2J_0(k_i\xi)}{k^3_i
J_1(k_i)}\nonumber\\
&& \hspace{-0.7cm}=\frac{\Phi_0\Theta_0^2\kappa}{2}
\left[1-\frac{Q^\ast}{2R M}-\left(1-\frac{Q^\ast}{2\eta R
M}\right)\frac{1}{\eta^3}\right](1-\xi^2)\nonumber \\
&&\hspace{-0.7cm}  =\frac{\Omega R^2 B_0\Theta_0^2\kappa}{2}
\left[1-\frac{1}{\eta^3}-\frac{Q^\ast}{2M
R}\left(1-\frac{1}{\eta^4}\right)\right](1-\xi^2)\ .\nonumber \\
\end{eqnarray}

Corresponding to this potential component of electric field
$E_{\|}$ will look like
\begin{eqnarray}
&&E_{\|}=-\frac{1}{R}
\frac{\partial\Phi}{\partial\eta}|_{\xi=constant} = \nonumber\\
&&\qquad-\frac{E_{vac}
\Theta^2_0\kappa}{2}\left(\frac{3}{\eta^4}-\frac{2Q^\ast}{M
R}\frac{1}{\eta^5}\right)(1-\xi^2)\ ,
\end{eqnarray}
where $E_{vac}\equiv(\Omega R/c)B_0$ is the characteristic
Newtonian value of the electric field generated near the surface
of a neutron star rotating in vacuum~\cite{d55}.

Using expression from the work of \cite{MusHar97} for the total
power carried away by relativistically moving particles
\begin{equation}
\label{Ldif} L_p=2(-c\int\rho\Phi\ dS)\ .
\end{equation}
one can obtain for the maximum of $L_p$
\begin{eqnarray}
&& (L_p)_{max}=\nonumber\\
&&\frac{3}{2}\kappa\left(1-\frac{Q^\ast}{2M
R}\right)\left[1-\kappa\left(1-\frac{Q^\ast}{2M
R}\right)\right]\dot{E}_{rot}\ ,\label{Lmax}
\end{eqnarray}
where
\begin{equation}
\dot{E}_{rot}\equiv\frac{1}{6}\frac{\Omega^4 B_0^2 R^6}{c^3
f^2(1)}=\frac{1}{f^2(1)}(\dot{E}_{rot})_{Newt}\
\end{equation}
and $(\dot{E}_{rot})_{Newt}$ is the standard Newtonian expression
for the magneto-dipole losses in flat space-time approximation.

One can see that obtained equation (\ref{Lmax}) differs from
corresponding equation from the work of \cite{MusHar97} for the
case of slowly rotating neutron star
$(L_p)_{max}=\frac{3}{2}\kappa(1-\kappa)\dot{E}_{rot}$ by
replacing $\kappa$ with $\kappa(1-Q^\ast/2MR)$.

%оценить малость добавки.  аков вывод - поправки пренебрежимы?

In terms of pulsar's period $P$ and its time derivative
$\dot{P}\equiv dP/dt$ equation (\ref{Lmax}) will look as

\begin{eqnarray}
&& (P\dot{P})_{max}=\frac{3}{4}\kappa\left(1-\frac{Q^\ast}{2M
R}\right)\nonumber\\ && \qquad
\left[1-\kappa\left(1-\frac{Q^\ast}{2M
R}\right)\right]\frac{I}{\tilde{I}}
\frac{1}{f^2(1)}(P\dot{P})_{Newt}\ ,\label{PP}
\end{eqnarray}
where
\begin{equation}
\label{LPP} (L_p)_{max}=-\tilde{I}(\Omega\dot{\Omega})_{max}
\end{equation}
and
\begin{equation}
(P\dot{P})_{Newt}\equiv\left(\frac{2\pi^2}{3c^3}\right)
\frac{R^6B^2_0}{I}\ .
\end{equation}
In equation (\ref{LPP}) $\tilde{I}$ is the general relativistic
moment of inertia of the star (see e.g.~\cite{r3})
\begin{equation}
\tilde{I}\equiv\int d^3\mathbf{x}\sqrt{\gamma}e^{-\Phi(r)}\rho
r^2\sin^2\theta\ ,
\end{equation}
where $e^{-\Phi(r)}\equiv 1/\sqrt{-g_{00}}$, $\rho(r)$ is the
total energy density, $\gamma$ is the determinant of the three
metric and $d^3\mathbf{x}$ is the coordinate volume element.

Expression (\ref{PP})  could be used to investigate the rotational
evolution of magnetized neutron stars with plasma magnetosphere. A
detailed investigation of general relativistic effects for
Schwarzschild stars in vacuum has already been performed by
\cite{pgz00}, who have paid special attention to the general
relativistic corrections that need to be included for a correct
modeling of the thermal evolution but also of the magnetic and
rotational evolution. It should be remarked, however, that in
their treatment \cite{pgz00} have taken into account the general
relativistic amplification of magnetic field  due to the curved
background spacetime, but did not include the corrections due to
the gravitational redshift. As a result, the general relativistic
electromagnetic luminosity estimated by \cite{pgz00} is smaller
than the one computed in paper \cite{r3} where all general
relativistic effects are taken into account. The evolution of
low-mass X-ray binaries hosting a neutron star and of millisecond
binary radio pulsars using numerical simulations that take into
account the detailed evolution of the companion star, of the
binary system, and of the neutron star has been studied in the
paper~\cite{lav05}.

%%%%%%%%%%%%%%%%%%%%%%%%%%%%%%%%%%%%%%%%%%
%
%
% Abdujabbarov
%
%
%%%%%%%%%%%%%%%%%%%%%%%%%%%%%%%%%%%%%%%%%%

\section{Linear Plasma Modes Along the Open Field Lines}

The theory of cascade generation of electron-positron plasma at
the polar cap region of a rotating plasma is developed by
\cite{rs75}. According to the theoretical model, because of the
escape of charge particles along the open field lines, a polar
potential gap is produced that continuously breaks down by forming
an electron-positron pair on a timescale of a few microseconds. A
photon of energy greater than $2mc^2$ produces an
electron-positron pair. The electric field of the gap accelerates
the positron out of the gap and accelerates the electron toward
the stellar surface. The electron moves along a curved magnetic
field line and radiates an energetic photon that goes on to
produce a pair as it has a sufficient component of momentum
perpendicular to the magnetic .eld. Recently, \cite{zhr97}
explained the $e-e^{+}$ pair production from a Crab-like pulsar.
Electrons and positrons are accelerated in opposite directions to
extremely high energies. The Lorentz factor $\gamma$ of the
"primary" electron and positron is given by
\begin{equation}
e\mathbf{E}\cdot \mathbf{B}c\approx \frac{e^2}{c^3}\gamma^4
\left(\frac{c^2}{r_c}\right)^2 ,
\end{equation}
where $r_c$ is the curvature radius of the local magnetic field
lines and $\mathbf{E}\cdot\mathbf{B}$ is the pulsar magnetospheric
electric field component along $\mathbf{B}$. For the Crab pulsar,
the above equation gives $\gamma$ of the order of $10^7$.
Curvature photons radiated by the primary electrons and positrons
have energy $\approx h\gamma^3 (c/r_c)$. Each primary electron and
positron would produce about $N_\gamma\approx (\gamma
e^2/hc)\approx 10^5$ curvature photons. To sustain the primary
current flow and account for the observed X-ray and $\gamma$-ray
luminosity from the Crab pulsar, the needed primary particle flux
is around $10^{33} s^{-1}$. This cascade of pair production,
acceleration of electrons and positrons along curved field lines,
curvature radiation, and pair production results in a "spark"
breakdown of the gap. Linear and nonlinear plasma modes near the
slowly rotating neutron star has been studied in the work of
\cite{ma00}. Here we will generalize the obtained results into the
case when the neutron star is considered in the braneworld.

Assuming a steady state thermodynamical equilibrium plasma state
in the polar cap region, we study the linear plasma modes along
the field lines. From the charge continuity equation
\begin{equation}
\frac{\partial\rho}{\partial t}+\left(\frac{\kappa}{r^3}-1 \right)
\Omega m\cdot \nabla\rho +\nabla\cdot (A \mathbf{j})=0\ ,
\end{equation}
the equation of motion of the charged particle
\begin{equation}
 \left[\frac{\partial}{\partial t}+
 \left(\frac{\kappa}{r^3}- 1\right) \Omega \cdot\nabla+ A
 \bf{v} \cdot \nabla  \right] \mathbf{p}=-q\nabla\Phi\ ,
\end{equation}
and the equation (\ref{Poiss}), one can easily derive the
following linearized equations:
\begin{eqnarray}
&& \hspace{-0.5cm}
   \nabla \left[\left(1-\frac{2M}{r}+\frac{Q^*}{r^2}\right)^{-1/2}
   \nabla \Phi\right]=-4\pi(\rho-\rho_{\rm GJ})\label{1steq}\ ,\\
&& \hspace{-0.5cm}
   \frac{\partial \rho}{\partial t'} + \nabla \frac{\partial}{\partial
   t'} \left(1-\frac{2M}{r}+\frac{Q^*}{r^2}\right)^{1/2}
   \mathbf{j}_{\rm s}\label{2ndeq}\ , \\
&& \hspace{-0.5cm}
   \frac{\partial \mathbf{v}_{\rm s}}{\partial
   t'}=-\frac{e_s}{m} \nabla \Phi\label{3rdeq}\ ,
\end{eqnarray}
where $\partial/\partial t'=\partial/\partial t+(\kappa /r^3-1)
\Omega \cdot \nabla$ is the global time derivative along ZAMO
trajectories, $\mathbf{j}_{\rm s}$ is the current density and
$e_{\rm s}$ is the electric charge of the particle,
Goldreich-Julian charge density $\rho_{\rm GJ}$ is defined by the
equation (\ref{rhoGJ}).

\begin{figure}
\includegraphics[width=0.45\textwidth]{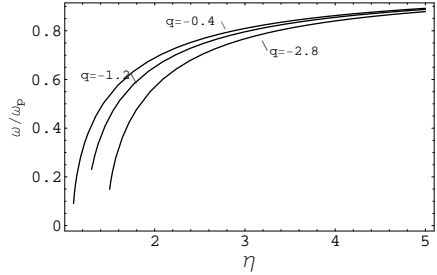}%

\caption{\label{fig:1stfig} The radial dependence of the $\omega$
for the different values of the brane tension $Q^*$ .}
\end{figure}

From the system of equations (\ref{1steq})--(\ref{3rdeq}) one can
get the following equivalent equation:
\begin{eqnarray}
&& \frac{\partial^2}{\partial t'^2}\left\{
\nabla\left[\left(1-\frac{2M}{r}+\frac{Q^*}{r^2}\right)^{-1/2}
   \nabla \Phi\right]\right\} + \nonumber \\ && \nabla \left[
   \omega_p^2\left(1-\frac{2M}{r}+\frac{Q^*}{r^2}\right)^{1/2}
   \nabla\Phi\right]=0\ ,  \label{difeqlin}
\end{eqnarray}
where $\omega_p^2=8 \pi\rho e_{\rm s}/m$. Using the following
notations
\begin{eqnarray}
&& \mathbf{E}=-\left(1-\frac{2M}{r}+\frac{Q^*}{r^2}\right)^{-1/2}
\nabla \Phi\ , \\ &&  \rho_{\rm GJ}=\frac{1}{4\pi}\nabla
\mathbf{E}_{\rm c}\ ,
\end{eqnarray}
equation (\ref{difeqlin}) can be rewritten as
\begin{equation}
\frac{\partial^2}{\partial t'^2}\nabla
\left(\mathbf{E}+\mathbf{E}_{\rm c} \right)+\nabla \omega_{\rm
p}^2 \left(1-\frac{2M}{r}+\frac{Q^*}{r^2}\right) \mathbf{E}=0\ .
\end{equation}
If we take into account that $\mathbf{E}_{\rm c}=const$, then we
have the following ordinary differential equation:
\begin{equation}
\frac{\partial^2 \mathbf{E}}{\partial t'^2} + \omega_{\rm p}^2
\left(1-\frac{2M}{r}+\frac{Q^*}{r^2}\right) \mathbf{E}
\end{equation}
which has solution of form
\begin{equation}
E=E_0 \exp\left(i \omega_p \sqrt{1-\frac{2M}{r}+\frac{Q^*}{r^2}}\
t\right)= E_0 \exp\left(i \omega t\right)\ .
\end{equation}
$\omega=\omega_p \sqrt{1-{2M}/{r}+{Q^*}/{r^2}}$ can be thought as
plasma frequency in general relativity, which is equivalent to the
gravitational redshift of the oscillation. Figure \ref{fig:1stfig}
shows the radial dependence of the frequency $\omega$ for the
different values of the brane parameter $Q^*$. The increase of the
module of the brane parameter causes shift of the plasma modes to
the red direction.

The global time derivative along ZAMO trajectories is defined as:
\begin{equation}
\frac{\partial}{\partial t'}=\frac{\partial}{\partial t}+ \left(
\frac{\kappa}{r^3}-1\right)\Omega\frac{\partial}{\partial \phi}\ ,
\end{equation}
that is one may define
\begin{equation}
{t'}={t}+ \frac{\phi}{\left({\kappa}/{r^3}-1\right)\Omega}\ ,
\end{equation}
and hence the solution of linear plasma mode is
\begin{eqnarray}
&& \hspace{-0.7cm}E=\nonumber \\
&& \hspace{-0.7cm} E_0 \exp\left\{i \omega_p
\sqrt{1-\frac{2M}{r}+\frac{Q^*}{r^2}}\ \left[{t}+
\frac{\phi}{\left({\kappa}/{r^3}-1\right)\Omega} \right]\right\}
.\label{efield}
\end{eqnarray}

Now, introducing new dimensionless parameters as $\epsilon=E/E_0$,
$\tau=\omega_p t$, $\chi=\omega_p/\Omega$, $\delta=\kappa/{8M^3}$,
and $q=Q^*/{M^2}$ we get from equation (\ref{efield}):
\begin{equation}\label{epsfield}
\epsilon=\sin\left[\sqrt{1-\frac{2}{\eta}+\frac{q}{\eta^2}}\
\left(\tau+\frac{\chi}{{8\delta}/{\eta^3}-1}\phi\right)\right]\ .
\end{equation}

\begin{figure}
\includegraphics[width=0.45\textwidth]{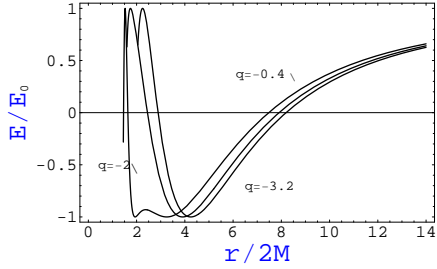}%

\caption{\label{fig:2ndfigs}  The radial dependence of the
dimensionless quantity $\epsilon$ for different values of the
dimensionless brane tension.}
\end{figure}

Figure \ref{fig:2ndfigs} shows the radial dependence of the
dimensionless quantity $\epsilon$ for different values of the
dimensionless brane tension. The intensity of the linear plasma
modes will be decreased with  the increase of the module of the
brane charge.

The both Fig. \ref{fig:1stfig} and Fig. \ref{fig:2ndfigs} show
that effect of the brane parameter is essential near the surface
of the Neutron star.

\section{Nonlinear Plasma Modes Along the Field Lines}

In this section we consider nonlinear plasma modes along the open
field lines around a rotating magnetized neutron star in
braneworld. The system of equations governing the nonlinear modes
can be written as
\begin{eqnarray}
&& \hspace{-0.3cm}
   \frac{\partial}{\partial
   t'}(n_s)+\nabla\cdot\left(
   \sqrt{1-\frac{2M}{r}+\frac{Q^*}{r^2}}n_s
   v_s\right)=0,\label{eq33}\\
&& \hspace{-0.3cm}
   \left(\frac{\partial}{\partial
   t'}+\sqrt{1-\frac{2M}{r}+\frac{Q^*}{r^2}} v_s \cdot
   \nabla\right)
   v_s=-\frac{e_s}{m}\nabla \Phi\ ,\label{eq34}\\
&& \hspace{-0.3cm}
   \nabla \cdot
   \left[\left(1-\frac{2M}{r}+
   \frac{Q^*}{r^2}\right)^{-\frac12}\nabla\Phi\right]=
   \nonumber
   \\
   && \ \ -4\pi\left(\sum_s n_s e_s - \rho_{GJ}\right). \label{eq35}
\end{eqnarray}
Now for simplicity one may consider
$v_s\cdot\nabla=v_{sr}\partial/\partial r$ (i.e., one-dimensional
wave propagation along $r$) and introduce a moving frame
$\tilde{\eta}=r-Vt'$, where $V$ is a constant. In the considered
moving frame, from equations (\ref{eq33}) and (\ref{eq34}) we get

\begin{eqnarray}
&& \hspace{-0.3cm} n_s=\frac{n_0 V}{V-(1-2M/r+Q^*/r^2)^{1/2}v_{s\tilde{\eta}}}\ , \label{eq36}\\
&& \hspace{-0.3cm} v_{s\tilde{\eta}}-\frac{1}{mV}\left(q_s \Phi
+\sqrt{1-\frac{2M}{r}+\frac{Q^*}{r^2}}\frac{e^2}{2mV^2}\Phi^2\right).
\label{eq37}
\end{eqnarray}

Using equations (\ref{eq36}) and (\ref{eq37}), in equation
(\ref{eq35}) we derive the nonlinear equation for the plasma mode
along the field line of the rotating neutron star in braneworld:
\begin{eqnarray}
&& 4\pi\sqrt{\tilde{A}}\rho_{GJ}=
%\nonumber \\&& \hspace{-0.7cm}
\frac{d^2\Phi}{\tilde{\eta}^2}-\frac{M}{\tilde{A} \tilde{\eta}^2}
\frac{d\Phi}{d\tilde{\eta}}+\nonumber
\\
&& \hspace{1.cm}\frac{\omega^2_{p0}\tilde{A}}{ V^2}
\frac{\Phi}{1-2\tilde{A} \left(\frac{e\Phi}{mV^2}\right)^2+\frac14
\tilde{A}^2 \left(\frac{e\Phi}{mV^2}\right)^4 }\ ,
\label{eqbezdim}
\end{eqnarray}
where
$\tilde{A}=1-\frac{2M}{\tilde{\eta}}+\frac{Q^*}{\tilde{\eta}^2}$

 Now, introducing dimensionless quantities
\begin{eqnarray}
&&  \bar{\Phi}= \frac{e\Phi}{mV^2}, \qquad
\bar{\eta}=\frac{\tilde{\eta}}{M}, \nonumber \\ &&
\bar{\omega}_{p0}=\frac{2M \omega_{p0}}{V}\ , \qquad \bar{A}= 1-
\frac{1}{\bar{\eta}}+\frac{q}{\bar{\eta}^2}
\end{eqnarray}
one can easily rewrite the equation (\ref{eqbezdim})
\begin{eqnarray}
&&\frac{d^2\bar{\Phi}}{d\bar{\eta}^2} +
\omega_{p0}^2\bar{A}\frac{\bar{\Phi}}{1-2\bar{A}
\bar{\Phi}^2+1/4\bar{A}^2 \bar{\Phi}^4}\nonumber\\&& \qquad
-\frac{1}{\bar{\eta}^{2}} A^{-1} \frac{d\bar{\Phi}}{d\bar{\eta}}=
\frac{\pi e}{mV^2M^2}A^{-1/2} \rho_{GJ}\label{numeric1}
\end{eqnarray}

\begin{figure}
\includegraphics[width=0.45\textwidth]{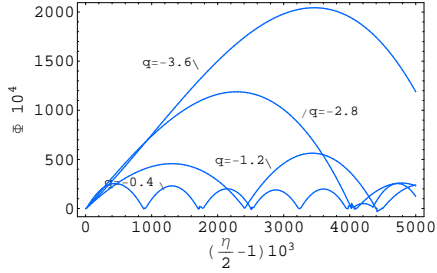}%

\caption{\label{fig:numeric} Nonlinear plasma modes in branworld
scenario; propagation of the nonlinear plasma modes near the
surface of the neutron star.}
\end{figure}

We numerically solve equation (\ref{numeric1}) in the polar cap
region ($\theta\simeq0$) of a neutron star subject to the
appropriate boundary conditions. Following Goldreich \& Julian
(1969) and Muslimov \& Harding (1997), we assume that the surface
of a polar cap and that formed by the last open field lines can be
treated as electric equipotentials. We therefore adopt the
condition $\Phi(r=R)=0$. Second, we require that the steady state
component of the electric field parallel to the magnetic field
vanishes at the polar cap surface, i.e.,
$d\Phi(r=R)/d\tilde{\eta}=0$. The boundary conditions can be
written as $\Phi(2)=\Phi'(2)=0$. The solution of equation
(\ref{numeric1}) for the different values of brane tension with
the mentioned boundary condition is shown graphically in Fig.
\ref{fig:numeric}. It is shown  that the Goldreich-Julian charge
density creates the initial potential on the surface. Near the
radius the potential is enhanced, and it propagates almost without
damping along the field lines in the presence of the brane charge.
The increasing the module of the brane charge causes to increasing
the amplitude of the nonlinear plasma modes.

\section{Charged Particle Acceleration in the Polar Cap of A
Slowly Rotating Star in Branewold} \label{acceleration}

The origin of radio emission from the polar cap is one of the most
mysterious of the remaining unsolved questions in the physics of
pulsars. One likely scenario is that particles are accelerated
along open magnetic field lines and emit $\gamma$-rays that
subsequently convert into electron-positron pairs in the strong
magnetic field. The combination of the primary beam and the pair
plasma then provides the mechanism of radio emission. For this
reason, it is interesting to study particle acceleration
conditions and the equations of motion in the pulsar magnetosphere
in braneworld models.

In the previous sections the effects of brane parameter on plasma
modes along the open field lines of a rotating, magnetized neutron
star are studied. In the paper \cite{ss03} the detail analyze of
the motion of the charged particle for a rotating pulsar polar cap
has been considered. Now we investigate the charged particles
acceleration in the region just above the polar cap surface of a
neutron star with brane charge.

Charges $e$ of the same sign as $\rho_{GJ}$ are extracted from the
polar cap and carry current $j$. We assume that this
charge-separated flow is freely supplied by the star with initial
velocity $v\ll c$ and neglect a binding energy of charges at the
surface. We will look for a steady state. All particles are in the
ground Landau state and flow along the field lines. The flow is
governed by the electric field $E_{\|}$ (parallel to $B$),
\begin{equation}
\frac{dp}{dt}=\frac{eE_{\|}}{mc}\ , \label{dpdt}
\end{equation}
where the momentum of the particle is given in units of $mc$. And,
for the parallel electric field $E_{\|}$ one can obtain the
equation
\begin{equation}
\nabla\cdot\mathbf{E}=\frac{dE_{\|}}{dz}=4\pi(\rho-\rho_{GJ}),
\qquad z\ll r_{\rm pc} \label{eqno44}
\end{equation}
where $r_{\rm pc}$ is the radius of the polar cap of the star and
$z$ is the distance from the stellar surface. Rewriting equation
(\ref{eqno44}) in terms of $d/dt = v d/dz$, $\tilde{a} =
j/cj_{GJ}$, $ \rho= j/v$, and $v = cp(1 + p^2)^{-1/2}$, one can
obtain
\begin{equation}
\frac{dE_{\|}}{dt}=4\pi j\left( 1-
\frac{\tilde{a}p}{\sqrt{1+p^2}}\right)\label{depdt}
\end{equation}
(see \cite{Bel07}), where
\begin{equation}
\tilde{a}(z)=\tilde{a_0}\frac{1-\kappa\left(1-{Q^*}/{2zM}\right)}{1
-\kappa\left(1-{Q^*}/{2zM}\right)\left( 1+z/R\right)^{-3}}.
\label{eqfora}
\end{equation}
Note that to obtain the expression for $\tilde{a}(z)$
(\ref{eqfora}) we have used the approximate value of the
Goldreich-Julian charge density at the polar cap region as
\begin{equation}
\rho_{GJ}=\frac{1}{2\pi}\frac{B\Omega}{N}\left(
1-\frac{\omega}{\Omega}+\frac{\omega}{\Omega}\frac{Q^*}{2zM}\right)\
.
\end{equation}
\begin{figure}
\includegraphics[width=0.45\textwidth]{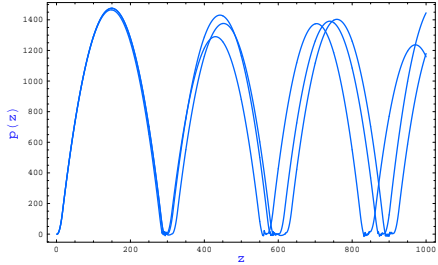}%

\caption{\label{fig:polacc} Dependence of the momentum of a
charged particle, extracted from polar cap for different value of
the brane parameter.}
\end{figure}

Finally the system of the equations (\ref{dpdt}), (\ref{depdt}),
and (\ref{eqfora}) is solved numerically.
 In Fig. \ref{fig:polacc} the dependence of the
momentum of a charged particle, extracted from polar cap for
different value of the brane parameter are shown. The values of
the parameters $B=3\times10^{12}G$, $P=1 s$, $\kappa=0.15$, and
$\tilde{a_0}=0.999$ are taken from the paper \cite{makApJ}. This
dependence shows that the  brane charge changes the period of the
oscillations.

\section{Conclusion}
\label{concl}

We have considered astrophysical processes in the polar cap of
pulsar magnetosphere in space-time of slowly rotating star in the
braneworld. In particular, the corrections caused by the brane
tension on the Goldreich-Julian charge density, electrostatic
scalar potential and accelerating component of electric field
being parallel to magnetic field lines in the polar cap region are
found. The presence of brane tension slightly modulates
Goldreich-Julian charge density near the surface of the star and
gives very important additional contribution to the generation of
accelerating electric field component in the magnetosphere near
the surface of the neutron star.

These results are applied to find an expression for
electromagnetic energy losses along the open magnetic field lines
of the slowly rotating star in the braneworld. It is found that in
the case of non-vanishing brane tension an important contribution
to the standard magneto-dipole energy losses expression appears.
Comparison of effect of brane tension with the already known
effects shows that it can not be neglected. Obtained new
dependence may be combined with astrophysical data on pulsar
periods slow down and be useful in further investigations on
possible detection of the brane tension.

It is also shown that the presence of brane tension has influence
on the conditions of particles motion in the polar cap region.
From derived results it can be seen, that brane tension modulates
the period of oscillations of particle's momentum. Obtained
numerical results can be useful in detecting or getting upper
limit  for  brane  parameter by comparing them with astrophysical
data on pulsar radiation. As further extension of this research
one may consider the propagation of the low-frequency
electrostatic wave in a nonuniform electronЦpositron pair
magnetoplasma of rotating neutron star containing density,
velocity, temperature and magnetic field
inhomogeneities~\cite{shukla08}.

\section*{Acknowledgments}

This research is supported in part by the UzFFR (projects 1-10 and
11-10) and projects FA-F2-F079, FA-F2-F061 of the UzAS and by the
ICTP through the OEA-PRJ-29 project.  BJA acknowledges the partial
financial support from the German Academic Exchange Service DAAD
and the IAU C46-PG-EA programs.

\end{document}